\documentclass[aps,prb,twocolumn]{revtex4}

\usepackage{amsmath}
\usepackage{color,xcolor}
\usepackage{graphicx}
\usepackage{hyperref}
\usepackage{bm}
\usepackage{amsfonts}
\usepackage{soul}

\def\ket#1{\left|#1\right\rangle}
\def\braket#1#2{\left\langle #1\right|\left.#2\right\rangle}

\def\be{\begin{equation}}       \def\ee{\end{equation}}
\def\bea{\begin{eqnarray}}      \def\eea{\end{eqnarray}}
\def\ba{\begin{array} }
\def\ea{\end{array} }
\def\bnum{\begin{enumerate} }
\def\enum{\end{enumerate}}

\def\=>{\Rightarrow}
\def\>{\rightarrow}

\def\eye2{Fathbb{I}}

\def\r{\vec{r}}
\def\d0{\Delta_{0}}

\usepackage{amsmath}

\begin{document}
\title{\bf Floating topological phases}
\author{Trithep Devakul}
\affiliation{Department of Physics, Princeton University, Princeton, NJ 08544, USA}
\author{S. L. Sondhi}
\affiliation{Department of Physics, Princeton University, Princeton, NJ 08544, USA}
\author{S. A. Kivelson}
\affiliation{Department of Physics, Stanford University, Stanford, CA 94305, USA}
\author{Erez Berg}
\affiliation{Department of Physics, Weizmann Institute of Science, Rehovot, Israel}

\date{\today }

\begin{abstract}
While quasi-two-dimensional (layered) materials can be highly anisotropic, their  asymptotic long-distance behavior generally reflects the properties of a fully three dimensional phase of matter.  However, certain topologically ordered quantum phases with an emergent 2+1 dimensional gauge symmetry 
can be asymptotically impervious to interplane couplings.
We discuss the stability of such ``floating topological phases,'' 
as well as their diagnosis by means of a non-local 
order parameter.
Such a phase can produce a divergent ratio  $\rho_{\perp}/\rho_{\parallel}$ of the inter-layer to intra-layer resistivity as $T\to 0$, even in an insulator where both  $\rho_{\perp}$ and $\rho_\parallel$ individually diverge.  Experimental observation of such a divergence  would constitute proof of the existence of a topological (e.g. spin liquid) phase.
\end{abstract}

\maketitle

\section{Introduction}
Although bulk materials are - obviously - three dimensional, it is common to encounter materials with highly anisotropoic electronic structure -  either quasi-1D  or quasi 2D structures.  Naturally, macroscopic thermodynamic correlation functions and linear  response properties of such materials are highly anisotropic, reflecting the microscopic anisotropy.  What is less obvious - but an intriguing possibility - is that such anisotropic systems might exhibit one form or another of a ``floating phase,'' i.e. a phase that exhibits asymptotic properties characteristic of a lower dimensional system~\cite{efetov,larkin,horowitz,golubovic,ohernlubensky,ohernlubenskytoner,
emerysliding,vishwanathsliding,sondhi,mukhopadhyay,tewari,tewari2,shtengel}.  For instance, the ground-state of a one-dimensional electron gas (1DEG) generically exhibits power-law  correlations  (e.g. Luttinger liquid or  Luther-Emery liquid behavior\cite{emery,solyom}); a quantum floating phase of a quasi 1D array of weakly  coupled wires, were such a phase to exist, would be a $T=0$ phase which exhibits power-law correlations along the wires, but exponentially falling correlations in the transverse directions.\cite{emerysliding,
vishwanathsliding,mukhopadhyay}  A classical ($T>0$) floating phase that has been considered\cite{golubovic,ohernlubensky,ohernlubenskytoner,horowitz} 
in a quasi-2D material of stacked planes, would exhibit power-law charge-density-wave (CDW) or superconducting (SC) correlations in-plane that fall exponentially with distance perpendicular to the planes.  

Disappointingly, for the most part whenever a lower dimensional system exhibits power-law correlations, arbitrarily weak higher dimensional couplings cause the onset of long-range order rather than a floating phase.  This reflects the fact that power-law phases are often critical in the sense that the susceptibility to one or another form of broken symmetry is infinite.  Even in more mundane phases - either disordered  or long-range ordered - in which the connected correlation functions fall exponentially with distance, the correlation length characterizing that falloff can be anisotropic reflecting the microscopic anisotropy of the system, but it is ultimately governed by a single 3d Ornstein-Zernike form at long enough distances.

There do appear to be special extreme circumstances in which such floating phases may be stable phases of matter~\cite{nayak}.  In most cases, however, the best one can hope for in the standard phases of matter is to find circumstances in which the simplest (and largest) higher dimensional couplings are in some way frustrated and rendered irrelevant, so that 3+1D behavior is only apparent beyond some extremely long (emergent) length scale.

The situation is more promising for topological phases.   Consider a stack of initially decoupled planes, each characterized by some form of topological order.  If we  were to turn on  interplane  couplings, this can effect all sorts of local correlations, but cannot effect the topological order.  Thus, as we will discuss (
formalizing and extending ground-breaking work of Senthil and Fisher\cite{SF}), if the couplings are weak enough, they cannot 
couple the topological order associated with each plane.  The topologically non-trivial properties of the state in the absence of interplane coupling - and in particular the fact  that there is an emergent gauge symmetry associated with each plane individually - survive in 3d, provided the microscopic couplings are sufficiently anisotropic.  In short, ``floating topological phases'' (as defined below) 
are predicted to be a generic feature of topological phases when (and if) they exist in 
the lower dimensional context.

From a more direct empirical perspective, a standard measure of the macroscopic anisotropy of a  system  is the  ratio  of transport coefficients in various directions, for instance, the ratio  of  the values of the resistivity tensor, $\rho_{\alpha\beta}$, along the different principle axes of  a  crystal.  In all conventional phases of matter, including a band insulator, a diffusive metal, and a disordered insulator (in the regime where it exhibits variable-range-hopping), even when the resistivity tensor is highly anisotropic, the ratio $\rho_{cc}/\rho_{aa}$ approaches a finite constant as $T\to 0$.  (Here and henceforth we take $c$ to  be the most resistive direction and $a$ the least.)
In contrast, in a floating phase, 
when the in-plane and interplane currents are carried by different combinations of elementary excitations, this ratio can diverge as $T\to 0$.  Note that this distinction can  apply in an insulator;  even when both $\rho_{aa}$ and $\rho_{cc}$ diverge as $T\to 0$, it is still possible to distinguish cases in which their ratio approaches a constant (as in conventional anisotropic phases of matter), or diverges. As was pointed out previously,\cite{werman} divergent anisotropies in thermal transport are also expected, although a typically large phonon contribution may mask the effect.

A diverging anisotropy ratio would certainly require some form of dynamical dimensional decoupling for which a more or less exotic explanation would be needed.  In the early days of cuprate high temperature superconductivity, Anderson and Zou\cite{anderson} proposed that just such a divergent resistivity anisotropy should be expected based on ideas 
of spin-charge separation.  Experiments carried out at high magnetic fields (to suppress the superconductivity) indeed were  initially  suggestive that this indeed occurs.\cite{boebinger}  More recently,\cite{li,pdw} in the stripe-ordered cuprate, LBCO, a form of dynamical layer decoupling has been  observed, in which the in-plane  resistivity, $\rho_{aa}$, drops rapidly below a well-defined onset temperature, $T_{onset}$, becoming immeasurably small below an apparent in-plane superconducting transition temperature, $T_{2d}$, while $\rho_{cc}$, the resistivity in the out-of plane direction, remains large and only vanishes at a substantially lower critical temperature, $T_{3d} < T_{2d}$.  The resistivity anisotropy increases by at least 3 orders of magnitude between $T_{onset}$ and $T_{2d}$, and is equal to  infinity within experimental error for $T_{2d} > T > T_{3d} $.  
For reasons that we will review and expand upon, it was suggested that a search for a related  divergent anisotropy -  in this case of the thermal conductivity tensor - could be used as a clear experimental signature of the existence of certain kinds of spin-liquid phases.  Similar ideas  were explored in the context of stacked quantum Hall layers by Balents and Fisher~\cite{balentsfisher}, Naud et al.~\cite{naud,naud2}, and
 Levin and Fisher~\cite{levinfisher}.

In this paper we offer a precise theoretical definition of a   topological floating phase.  We propose that a form of absolute stability - even in the presence of (weak) higher dimensional couplings -- may   be considered as a defining feature of  certain sorts of spin liquid phases.  
 Floating topological phases are also closely related to fracton topological ordered phases~\cite{fractonchamon,fractonhaah,fractonyoshida,fractonvijay}, which have attracted much theoretical attention in recent years~\cite{nandkishorereview,pretkoreview}.
Fracton topologically ordered phases host fractionalized quasiparticles with restricted mobility and a subextensive ground state degeneracy, all properties shared by floating topological phases.
The phases studied in this paper are therefore very basic versions of fracton phases.

This paper is organized as follows.
We define and discuss gapped floating topological phase in Sec~\ref{sec:gapped}, as well as their  their diagnostic in terms of non-local correlation functions in Sec~\ref{sec:fmop}.
In Sec~\ref{sec:gapless}, we extend this analysis to various gapless floating topological phases, and discuss their stability to interlayer couplings.
In Sec~\ref{sec:experimental}, we demonstrate that a diverging conductivity ratio may serve as an experimentally accessible signature of a floating topological phase.
Finally, we end with some concluding remarks in Sec~\ref{sec:conclusion}.

\section{Gapped topological floating phases}\label{sec:gapped}

We begin with the simplest possible case: a floating phase of gapped  topological orders in two spatial dimensions (i.e. 2+1D including the time dimension).
In such models, topological properties such as the ground state degeneracy and statistical properties of fractionalized quasiparticle excitations are stable to arbitrary interlayer couplings.
As the models we will consider later take the form of stacked gauge theories coupled to matter, 
we will first begin with a simplest such example: the stack of 2+1D Ising gauge theories coupled to Ising matter.
This model is equivalent to a stack of 2+1D Kitaev toric codes.\cite{kitaev}

The 2+1D Ising gauge theory (IGT) on the square lattice is described~\cite{kogut,fradkinshenker} 
by a model with matter degrees of freedom $\tau_r$ living on the sites $r$, and gauge degrees of freedom $\sigma_{\ell}$ living on the links $\ell=(rr^\prime)$ connecting nearest neighbor sites $r$ and $r^\prime$. 
The IGT Hamiltonian is 
\begin{eqnarray}
\begin{split}
H_{IGT} =& -K\sum_{\square}\prod_{ \ell  \in \square} \sigma_{\ell}^z 
- \Gamma \sum_{\ell}\sigma_{\ell}^x\\
 &- J\sum_{ 
 (rr^\prime)} 
 \tau^z_r \sigma_{
 ( rr^\prime)}^z\tau^z_{r^\prime} 
 - \Gamma_M \sum_{r} \tau^x_r
 \end{split}
\end{eqnarray}
where the first sum is over square plaquettes, and
$\sigma^{x,y,z}$ and $\tau^{x,y,z}$ are Pauli matrices acting on the $\sigma$ and $\tau$ degrees of freedom respectively.
Here the first  two terms are the ``Maxwell'' terms, and the remaining terms are the gauge-invariant matter terms.
This Hamiltonian is invariant under local gauge transformations, $G_r^\dagger H G_r = H$, generated by 
\begin{equation}
G_r = \tau^x_r \prod_{r^\prime} \sigma^x_{(r r^\prime)}
\end{equation}
on each site $r$, where the product is over the four nearest neighbors.
We take the physical subspace to be the one with $G_r=1$ on every site.
$K>0$ favors a zero-flux ground state, and $\Gamma$ makes the gauge field dynamical.

This model describes a deconfined $\mathbb{Z}_2$ gauge theory when $K/\Gamma \gg 1$ and $\Gamma_M/J\gg 1$.
In this limit, $H_{IGT}$ describes a perturbed version of Kitaev's toric-code model~\cite{kitaev} --- this can be seen by 
working in the gauge $\tau_r^z=1$, 
in which $H_{IGT}$ 
can be expressed entirely in terms of the gauge-fields
\begin{eqnarray}
\begin{split}
\tilde H_{IGT}
=& -K\sum_\square \prod \sigma^z - \Gamma_M\sum_{+} \prod \sigma^x \\
& - J \sum_{\ell}\sigma^z_{\ell}  - \Gamma \sum_{\ell}\sigma^x_{\ell}\\
=& H_{TC} - J\sum_{\ell} \sigma^z_{\ell} - \Gamma \sum_{\ell}\sigma^x_{\ell}
\end{split}
\label{eq:igtgaugefix}
\end{eqnarray}
where the $K$ and $\Gamma_M$ terms are 
the stabilizers of the toric code Hamiltonian $H_{TC}$, and $J$ and $\Gamma$ are small $\sigma^z$ and $\sigma^x$ perturbations that make the model dynamical.

The Toric Code Hamiltonian $H_{TC}$ possesses non-trivial topological order which is stable to arbitrary local perturbations~\cite{bravyi}.
When placed on a torus, $H_{TC}$ has four exactly degenerate ground states 
in the thermodynamic limit.
These ground states may be distinguished via non-local Wilson and 't Hooft operators.
Define the Wilson loop operator 
\begin{equation}
W_C = \prod_{l\in C} \sigma_l^z
\label{Wilson}
\end{equation}
where $C$ denotes a closed loop on the square lattice, and $l\in C$ are all the links involved.
We may also define the dual Wilson loop (or 't Hooft operator),
\begin{equation}
V_{\overline{C}} = \prod_{l\in \overline{C}} \sigma_l^x
\label{Hooft}
\end{equation}
where $\overline{C}$ denotes a loop on the dual square lattice, and $l\in\overline{C}$ are all the links cut by $\overline{C}$.
These operators commute with $H_{TC}$.
Let $W_1$ and $V_1$ denote the non-contractible Wilson loops going around the torus in the $x$ direction, and similarly $W_2$ and $V_2$ along $y$.
The operators $(W_1,V_2)$ and $(W_2, V_1)$ generate Pauli algebras on the 4-dimensional ground state manifold.
The ground state degeneracy is stable to $J$ and $\Gamma$ perturbations 
due to the fact that these non-contractible Wilson loop operators only appear at high order $\mathcal{O}(L)$ in perturbation theory where $L$  is the  circumference  of the torus; 
any lifting of the degeneracy is thus exponentially suppressed at large $L$, going as $\sim(J/\Gamma_M)^L$ or $\sim(\Gamma/K)^L$.

Next, consider the bilayer of two such 
systems, with some weak coupling between them
\begin{equation}
H = H_{IGT}^{(1)} + H_{IGT}^{(2)} + \lambda H_{inter}
\label{eq:bilayerham}
\end{equation}
where $H_{inter}$ contains local terms coupling the two layers.
On a torus, this bilayer now has a $4^2$-fold degenerate ground state manifold, which is only split perturbatively by the interlayer couplings at order $\lambda^L$ 
as before.
This simply describes a new topological order, which inherits all its topological properties from the stack of two decoupled toric codes.
Indeed, this is simply the $\mathbb{Z}_2\times\mathbb{Z}_2$ generalization of the toric code (which describes the gauge theory of a bilayer Ising model in which each layer has a separate $\mathbb{Z}_2$ symmetry).

Now, let us consider a 3+1D system on a 3-torus obtained by stacking $L$ such models in the $xy$ plane along the $z$ direction, allowing for small (but arbitrary) local perturbations. 
This model will have a robust $4^L$ ground state degeneracy which is stable to any arbitrary small interlayer interactions.
{ We define a $D+1$ dimensional system to be in a non-trivial gapped floating topological phase if it can be smoothly connected (via a finite depth local unitary transformation) to a decoupled stack of $d+1$ dimensional topologically ordered systems, where $0<d<D$.}
In the cases we consider, $D=3$ and $d=2$.
The stack of toric code models, with weak interlayer coupling terms, realizes a non-trivial floating topological phase by this definition.

Quasiparticle excitations of this model are constrained to move within a single 2+1D $xy$ plane.
As an aside, we note that decoupled stacks of topologically ordered planes, exactly as we have formulated, have appeared multiple times~\cite{slaglekim,shirleyslaglechen} in the literature of fracton topological order~\cite{nandkishorereview,pretkoreview}.  
If we take the definition of a fracton topological order to be a subextensive ground state degeneracy $\ln GSD\sim L$ on a $3$-torus and subdimensional quasiparticle excitations, then the stack of 2+1D topological orders are indeed (very simple versions of) fracton models.
These may also be obtained as gauge theories of models with planar subsystem symmetries along each $xy$ plane (a 1-foliated planar subsystem symmetry~\cite{devakulshirleywang}).

The low energy field theory description of the single-layer toric code is given by the 2+1D BF theory, or equivalently the Chern-Simons theory~\cite{hansson,freedman}
\begin{equation}
\mathcal{L}_{CS} = \frac{K_{IJ}}{4\pi}\epsilon^{\mu \nu \rho} a_\mu^I \partial_\nu a_\rho^J
\end{equation}
with the $2\times 2$ $\mathbf{K}$ matrix $\mathbf{K}^{TC}=2\sigma^x$.
The ground state degeneracy of such a model on a manifold of genus $g$ is then given by $GSD = |\det \mathbf{K}|^g$.
The bilayer toric code described by the Hamiltonian Eq~\ref{eq:bilayerham}
then admits a similar low-energy description, except with the $4\times 4$ $\mathbf{K}$ matrix given by the direct sum $\mathbf{K}^{2TC} = \mathbf{K}^{TC}\oplus \mathbf{K}^{TC}$.
The floating topological phase of toric codes discussed above may then be characterized by the extensively large $\mathbf{K}$ matrix $\mathbf{K}^{float}=\mathbf{K}^{TC} \oplus \dots \oplus \mathbf{K}^{TC}$ (such ``giant'' K matrices also appear in the classification of fracton phases~\cite{shirleyslaglechen2}).
This can be generalized to stacks of general Abelian topological phases characterized by the matrix $\mathbf{K}$~\cite{wenzee}. 

We also note the interesting possibility of offdiagonal elements in the large $\mathbf{K}$ matrix which couples different layers.
These types of systems have been studied~\cite{qiu,qiu2,naud,naud2} and found to exhibit interesting behavior (such as an irrational braiding statistic) which cannot be found in 2+1D systems.
Such systems are not floating topological phases by our definition as they cannot be deformed to the decoupled limit, but nevertheless have an emergent decoupled gauge symmetry.

Finally, most of our discussion can also be extended to stacks of non-Abelian topological orders~\footnote{More specifically, we concern ourselves with only \emph{non-invertible} topological orders, which possess fractionalized quasiparticles.
Stacks of invertible topological phases~\cite{invertible} need not form stable floating phases.}.  
Such a phase can be characterized by the topological properties of the quasiparticle excitations, such as the fusion coefficients and topological spin~\cite{wenclass}.
Stacking two layers results in a new phase whose quasiparticles are directly inherited from the individual layers (see Ref~\onlinecite{wenclass}), and survive interlayer couplings as long as the gap is not closed.
The floating phase of gapped non-Abelian topological orders are therefore also stable.

\section{Floating phases via the 
Fredenhagen-Marcu order parameter}\label{sec:fmop}

The goal of this section is to 
differentiate a floating topological phase from either the 3+1D topological order or the trivial phase by means of a correlation function.

\subsection{Usual deconfinement diagnostic}
We first review how this is done in the usual case of diagnosing deconfinement in the Ising gauge theory~\cite{gregor}.
Let us take $W(L)$
to be the Wilson loop defined in Eq. \ref{Wilson} along the contour $C$ taken to be an $L\times L$ square. 
For a pure gauge theory without dynamical matter ($J=0$ in Eq~\ref{eq:igtgaugefix}), the scaling of the expectation value of the Wilson loop is sufficient to diagnose deconfinement:  for large $L$, $\ln \langle W(L) \rangle \sim -L$ scales linearly with the perimeter of the loop in the deconfined phase, but in the confined phase scales with the area, $\ln \langle W(L) \rangle \sim -L^2$.
However, as soon as $J\neq 0$, the Wilson loop scales with the perimeter in both the deconfined and confined phases, and therefore fails to distinguish between the two.

To correct this shortcoming, 
consider (an equal time formulation of) the Fredenhagen-Marcu order parameter~\cite{gregor,fredenhagen} (FMOP).
Define 
\be
W_{\frac 12}(L)\equiv \tau^z_r\tau^z_{r^\prime}\prod_{l\in C_{\frac 1 2}} \sigma^z_l
\label{horseshoe}
\ee
to be a gauge-invariant open Wilson line (or horseshoe) operator, where $C_{\frac 1 2}$ is the $L\times L/2$ horseshoe terminated at sites $r$, $r^\prime$, obtained by cutting $C$ in half.  
Similarly, let $W_{-\frac{1}{2}}$ be the other half of the Wilson loop.
We note that a more general geometry is possible, we simply choose to horseshoe shape for simplicity.
It can be shown (see below) that the ratio 
\be
R(L) = \langle W_{\frac 12}(L) \rangle \langle W_{-\frac{1}{2}}\rangle / {\langle W(L) \rangle}
\label{RofL}
\ee
goes to $0$ in the deconfined phase, 
while $\lim_{L\to \infty} R(L) = R_0 >0$ in the confined phase.  
Thus, other than in the special case $J=0$, this provides a suitable signature of a deconfined phase. 
 
This behavior can be understood in many ways.  
If we adopt the same gauge choice as in Eq. \ref{eq:igtgaugefix}, we can express the ground-state of the Hamiltonian in the $\sigma^x$ basis as
\be
\ket{\psi} = \sum_c \alpha_c \ket{c},
\ee
 where $c=\{\sigma^x_l\}$ label all the configurations.
The ground-state of the toric code is 
the equal amplitude superposition of all configurations of closed $\sigma^x=-1$ loops, with zero amplitude for all other $c$;  it is  a loop condensate.  
With perturbations, the weights of each configuration in the ground state are no longer exactly equal and configurations with open strings now exist, ableit with weights that are exponentially small in the separation between the two endpoints $\ell$ going as $\sim (J/\Gamma_M)^\ell$.

The expectation value of the 
horseshoe operator (Eq. \ref{horseshoe})
\begin{equation}
\langle W_{\frac 1 2} \rangle = \sum_{c} \alpha_c^* \alpha_{c^\prime}
\end{equation}
where $\ket{c^\prime} = W_{\frac 1 2}\ket{c}$ is the configuration $c$ with $\sigma^x$ flipped along the support of $W_{\frac 1 2}$. (The terminal factors of $\tau_r^z$ in Eq. \ref{horseshoe} are  set equal to 1 by the choice of gauge.)
 There is an analogous expression for the expectation value of $W$. 
To see how these considerations distinguish the two phases, we use this expression to compute $R(L)$ at  points deep inside the respective phases.

Manifestly,  $\langle W_{\frac 1 2}(L)\rangle $ vanishes in the toric-code ground-state, since $W_{1/2}$ generates a string from $r$ to $r^\prime$, meaning that for any $c$ such that $\alpha_c \neq 0$, $\alpha_{c^\prime} =0$.  In the perturbed problem,
$\langle W_{\frac 1 2}\rangle$ is not identically zero, but decays exponentially with $L$. The form of its decay can be derived using perturbation theory in both $J$ and $\Gamma$: $\langle W_{\frac{1}{2}}(L) \rangle\sim e^{-(2 a + b)L}$. Here, $a\sim (\Gamma/K)$ comes from the sides of the horseshoe and $b\sim -\ln (J/\Gamma_M)$ comes from the string from $r$ to $r^\prime$. The Wilson loop scales as $\langle W(L)\rangle\sim e^{-4 a L}$. 
Thus $R(L)\sim e^{-2 b L}\rightarrow 0$ in the deconfined phase. 

To characterize the confined phase, consider the ground state of Eq. \ref{eq:igtgaugefix} in the large $\Gamma$ limit.  Here, the significant configuration are those which are  mostly polarized with $\sigma^x=+1$, plus small loop fluctuations (suppressed by factors of $(K/\Gamma)^A$ where $A$ is the enclosed area) and open line fluctuations (exponentially suppressed in their length as $(J/\Gamma)^\ell$) 
For large $L$, {$\langle W_{1/2}(L) \rangle \sim (J/\Gamma)^{2L}$ and $\langle W(L)\rangle \sim (J/\Gamma)^{4L}$.}
Hence, $R(L)\rightarrow \mathrm{const}$ in the large $L$ limit in the confined phase.
Some intuition can be gained along the special $\Gamma=0$ axis, where $R(L)$ is (the square of) the original Ising $\sigma^z \sigma^z$ correlation function, and the approach to a constant can be understood in terms of spontaneous symmetry breaking.

The generalization of this FMOP construction beyond $\mathbb{Z}_2$ is possible and discussed by Gregor et al~\cite{gregor}.

\subsection{Floating deconfinement diagnostic}
We now turn our discussion to the problem of diagnosing a floating phase of 2+1D topological orders from either 
a confining phase or a fully 3+1D topological ordered phase.
To do this we consider an anisotropic  3D generalization of the gauge-fixed version of the Ising gauge theory (Eq. \ref{eq:igtgaugefix}) with qubits  defined on the nearest-neighbor bonds of a tetragonal lattice, with couplings $K,J,\Gamma$ for the in-plane terms, and $K_\perp, J_\perp, \Gamma_\perp$ for those involving bonds in the interplane direction.
This system now has the same Hilbert space as that of the 3+1D toric code. 
Indeed, for $J=J_\perp=\Gamma=\Gamma_\perp=0$ and $K=K_\perp$, this model reduces to the 3+1D toric code, and to stacks of 2+1D toric codes if we then take the limit $\Gamma_\perp \rightarrow \infty$, $K_\perp\rightarrow 0$.

It is  thus clear that in different limits, this one model can support all three of the possible phases in question.
Our diagnostic for floating topological order is inspired by the usual deconfinement diagnostic just discussed.

Let $W
(z;L)$ be the 
Wilson loop 
on  a $L\times L$ loop in
the $z^{th}$ plane. 
We further split 
the loop into two equal  horseshoes such that $W(z;L) = W_{\frac 12}(z;L)\  W_{-\frac 12}(z;L)$.
where $W_{\frac 12}$ is defined on the left horseshoe and $ W_{-\frac 12}$  on the  right.
We then consider the ratio
\begin{equation}
R_2(L) = 
\frac{W_{\frac 12}(z;L)\  W_{-\frac 12}(z+1;L)}{\sqrt{W(z;L)\  W(z+1;L)}}.
\label{eq:equaltimeop}
\end{equation}
As we will see,  in analogy with the  previous analysis,
\begin{equation}
\lim_{L\rightarrow\infty} 
R_2(L)
=
\begin{cases}
0 \hspace{0.5cm}& \text{deconfined floating}\\
\mathrm{const} \hspace{0.5cm}& \text{confined or 3+1D deconfined}\\
\end{cases}
\end{equation}
distinguishes between the deconfined floating topological phase 
from a confined or fully 3+1D deconfined topological phase.

First, consider the decoupled limit with all the interplane couplings set equal to 0: 
now $R_2(L)$ factors as 
$R_2(L)=R(L)$, where $R(L)$ is 
the FMOP for 
a single 2+1 d plane defined in Eq. \ref{RofL}. Thus 
in this case, for the deconfined phase, $R_2(L)\rightarrow 0$, while for the confining phase $R_2(L)\rightarrow {\rm const} >0$.
It is easy to see that these results remain true even in the presence of arbitrary small perturbations.

Finally, consider the case of an isotropic 3+1D phase.  The numerator is now the expectation value of a Wilson loop minus only the two vertical bonds - the full Wilson loop  thus  differs from this by a factor which is order $O(1)$ (i.e. independent of $L$). Moreover, this Wilson loop  can be viewed as a slightly distorted relative of the  the Wilson loop in the denominator other 
except for two ``kinks'' where it changes over between planes $z$ and $z+1$.  
These kinks 
also make an $O(1)$ correction to the total expectation value of the Wilson loop.
The scaling of numerator and denominator again cancel, so independent of whether the 3+1D system is confining or not, $R_2(L)\rightarrow\mathrm{const}$.

In short, the vanishing of $R_2(L)$ at large $L$ is a signature of a topological floating phase.

Owing to the emergent subdimensional Lorentz symmetry of the floating topological phase, the order parameter may be oriented in various space-time directions (within the $x,y,\tau$ subspace), each of which have a physical interpretation~\cite{gregor}.
The order parameter Eq~\ref{eq:equaltimeop} corresponds to an equal time-slice orientation.  
We will now discuss another orientation of this order parameter (which in the usual case corresponded to the order parameter discovered by Fredenhagen and Marcu~\cite{fredenhagen}).
Let us denote by $|\psi_z\rangle$ a trial state with two (spinon) excitations located at $\vec{r}$ and $\vec{r^\prime}=\vec{r}+L\hat{x}$ on plane $r_z=z$, defined by
\begin{equation}
|\psi_z\rangle = \tau^z_r \tau^z_{r^\prime} V_{r r^\prime}(-T/2)\ket{\mathrm{GS}}
\end{equation}
where $V_{r r^\prime}(-T/2) \equiv e^{- H T/2} V_{r r^\prime} e^{HT/2}$, $V_{r r^\prime}=\prod_{l} \sigma^z_l$ is the (non gauge-invariant) Wilson line operator connecting the points $r$ and $r^\prime$, and $\ket{\mathrm{GS}}$ is the ground state.
$V(-T/2)$ acts on the ground state by creating two ``defects'' at $r$ and $r^\prime$ where $G_r=-1$, and (in the limit of large $T$) projects to the lowest energy state with such defects.  

The order parameter is given by
\begin{equation}
\tilde{R_2}(L,T) = \frac{\braket{\psi_z}{\psi_{z+1}}}
{\sqrt{
\braket{\psi_z}{\psi_z}
\braket{\psi_{z+1}}{\psi_{z+1}}
}}
\end{equation}\label{eq:equalxop}
which will exhibit the same asymptotic behavior as $L,T\rightarrow\infty$ as $F(L)$.
In this picture, we see that $\tilde{R_2}$ is probing the orthogonality of the trial spinon states on plane $z$ and $z+1$.  
In the deconfined floating phase, $\ket{\psi_z}$ and $\ket{\psi_{z+1}}$ will be orthogonal, since spinons cannot tunnel between planes.
However, in a fully 3+1D deconfined phase a spinon may move between the two planes and so $\braket{\psi_z}{\psi_{z+1}}\neq 0$.  
While $\tilde{R}$ is useful conceptually, in condensed matter systems where the gauge symmetry is emergent, the equal-time formulation $R_2$ (Eq~\ref{eq:equaltimeop}) should be used~\cite{gregor}.

Finally, all of this discussion can be extended beyond $\mathbb{Z}_2$.
One simply replaces $V$ with the appropriate Wilson line operator and $\tau$ by the appropriate charged matter operator for gauge invariance~\cite{gregor}.  

\section{Gapless floating topological phases}\label{sec:gapless}
In this section, we analyze the stability of gapless floating topological phases.
We first consider the floating phase of gapless Dirac matter coupled to a gapped $\mathbb{Z}_2$ gauge field.
We then go on to consider the floating phase of Dirac fermions coupled to a $U(1)$ gauge field, in which both the matter and gauge sectors are gapless.
We show that these gapless floating phases are stable to interlayer couplings in the renormalization group (RG) sense.
That is, the interlayer couplings are irrelevant perturbations: at long distances and low energies, the system flows back to the decoupled limit.

\subsection{Gapless matter, gapped gauge}
\label{sec:Z2Dirac}
Let us first consider the case with gapless Dirac matter coupled to a gapped gauge field. 
At low energies, a single 2+1D layer is described by Dirac fermions hopping in a background static gauge field. 
This describes, for example, the gapless phase of Kitaev's Honeycomb model~\cite{kitaevhoneycomb}, in which case the minimum energy configuration for the gauge field is equivalent to the trivial (flux-free) configuration.

The low-energy continuum Hamiltonian for a single layer $l$ is
\begin{equation}
H_l = \int d^2 r \psi^\dagger_l(\r) \left[-i v \vec{\sigma}\cdot \vec{\partial} \right] \psi_l(\r)
\label{eq:Z2Dirac}
\end{equation}
where $\psi^\dagger_l(\r)$ ($\psi_l(\r)$) is a 2-component spinor which creates (annihilates) a complex fermion at position $\r=(x,y)$ on layer $l$,  $v$ is the Fermi velocity, $\vec{\partial} = (\partial_x,\partial_y)$, and $\vec{\sigma}=(\sigma^x,\sigma^y)$ is a vector of Pauli matrices acting on the spinor indices~\footnote{This low-energy form is obtained from Kitaev's Honeycomb model by first combining the two Majorana cones into a single Dirac cone.}.

The corresponding Euclidean action for the single layer is  
\begin{equation}
S_l = \int d^2 r d \tau  \bar{\psi}_l(\r,\tau) \gamma^\mu \partial_\mu \psi_l(\r,\tau)
\end{equation}
where summation over $\mu=\tau,x,y$ is implied, $\gamma^\mu = (\sigma^z, \sigma^y, -\sigma^x)$, $\bar{\psi} = \psi^\dagger \sigma^z$, and we have rescaled coordinates to set $v=1$.

The gauge fields are fully gapped out, and hence do not appear in the low-energy description.
Nevertheless, they are important as they restrict the terms which are allowed to appear to only those which are gauge invariant.
When we have multiple layers, gauge invariance within each layer implies that any interlayer term must consist of operators which are individually gauge invariant on each layer.
Crucially, this forbids quadratic interlayer hopping terms, which are not gauge invariant within a single layer.
The simplest gauge-invariant terms coupling two layers are four-body interaction terms such as $(\bar\psi_l \psi_l)(\bar\psi_{l^\prime} \psi_{l^\prime})$ between two layers $l$, $l^\prime$.  
However, all such quartic terms are irrelevant at the $2+1$D Dirac fermion Gaussian fixed point with $S=\sum_l S_l$.  
This can be seen by simple power counting:
The field $\psi$ has length dimensions $[\psi]=L^{-1}$, and so the quartic term has dimensions $[(\bar{\psi} \psi)^2] = L^{-4}$.
Under an RG transformation in which we rescale time and the two continuous spatial dimensions, the quartic term therefore flows to zero.
If the system were to flow to a fully $3+1$D phase, we would instead expect that the interlayer couplings would increase under this RG flow. 
In this case, however, the system flows back to the decoupled layers limit at large distances.  
This system is therefore an example of a stable gapless floating phase.

Finally, we note that there are single-layer terms which are relevant.
These include quadratic terms of the form $\bar{\psi} \gamma^\mu \psi$, which may open up a gap or create a Fermi surface.  
In  assessing  the stability of  these phases, we have implicitly assumed that these terms are forbidden by symmetries of the microscopic Hamiltonian, and that the microscopic interlayer couplings also respect such symmetries.

\subsection{Gapless matter, gapless gauge}

Let us now consider the 
case of Dirac fermions coupled to a gapless $U(1)$ gauge field.  
The pure $U(1)$ gauge theory is confining in $2+1$D, but a stable deconfined phase can exist when coupled to a large number of Dirac fermions~\cite{hermele}.
In this situation, the low energy continuum description of each $2+1$D layer $l$ is simply large-$N$  quantum electrodynamics: $N$ Dirac fermion flavors $\psi_{i,l}$ ($i=1,\dots,N$) coupled to an emergent gauge field $a_{\mu,l}$.
The Euclidean Lagrangian is
\begin{equation}
\mathcal{L}_l^{\mathrm{QED}_3} =\sum_{i=1}^N \bar{\psi}_{i,l} \gamma^\mu (\partial_\mu + i a_{\mu,l}) \psi_{i,l} + \frac{1}{4 e^2} f_{\mu \nu,l} f^{\mu \nu}_l
\end{equation}
where $f_{\mu\nu,l} = \partial_\mu a_{\nu,l} - \partial_\nu a_{\mu,l}$ is the field strength. 
The gauge transformation sends $\psi_l \rightarrow e^{i \alpha_l} \psi_l$ and $a_{\mu,l}\rightarrow a_{\mu,l} - \partial_\mu \alpha_l$ for an arbitrary spacetime function $\alpha_l(\r,\tau)$ on each layer $l$.

The Maxwell term, although included, is irrelevant at large $N$.
This is exemplified by the fact that with a clever choice of non-local gauge fixing term, the gauge photon propagator can be written in such a way that the $e^2\rightarrow\infty$ limit can be taken at the beginning of a calculation~\cite{chesterpufu}.
Indeed, as written, we have $[a]=L^{-1}$ and so $[f_{\mu\nu} f^{\mu\nu}]=L^{-4}$ is irrelevant, while the coupling $[\bar{\psi} a \psi]=L^{-3}$ is marginal.

As in the previous case, the simplest interlayer coupling terms 
that are gauge-invariant are either quartic in the fermion operators or pure-gauge (of the form $f_{\mu\nu,l}f^{\mu\nu}_{l^\prime}$), both of which are strictly irrelevant perturbations at large $N$.
However,  as before, 
we have implicitly assumed that, for reasons of symmetry, 
relevant single-layer terms are not present.  
These now include, for example, a Chern-Simons term.

Like in the gapped case, rather than the low energy model with a separate gauge symmetry on each individual layer, a more natural starting point is a single anisotropic emergent 3+1D gauge field with interlayer couplings much weaker than intralayer.
In this case, one has a gauge fields $a_{z,l}$, and gauge-invariant interlayer hopping terms of the form $\bar{\psi}_{l+1} e^{i a_{z,l}} \psi_{l}$ are allowed.
In this limit, $a_{z,l}$ are strongly fluctuating (and therefore gapped), so they can be integrated out resulting in a local effective action with a separate gauge symmetry on each individual layer.
We go through this explicitly in Appendix~\ref{app:anisotropic_lgt} for an anisotropic $U(1)$ lattice gauge theory with fermions.
Such a system is partially confined: fractionalized quasiparticles are confined along $z$, but deconfined within each layer.

Such ``layered'' phases of the $U(1)$ gauge theory have been studied previously~\cite{funielson,hulsebos} and found to be stable in higher than $3+1$D.
These are examples of stable higher-dimensional gapless floating topological phases.

\subsection{Gapless matter with a Fermi surface}

Finally, we turn to the case where the emergent (gauged) fermions form a Fermi surface with a finite density of states.\cite{Anderson1,BZA} This situation may arise in a layered, gapless spin liquid. We first consider the simpler case of a $\mathbb{Z}_2$ gauge field, where the gauge degrees of freedom are gapped, and then comment on the more complex case of a U(1) gapless gauge field. 

\subsubsection{Fermi surface with $\mathbb{Z}_2$ gauge field}

In the case of gapless fermions coupled to a $\mathbb{Z}_2$ gauge field, the only low-energy degrees of freedom are the fermions. At a single layer $l$, the effective Hamiltonian is
\begin{equation}
    H_{\mathbb{Z}_2\,FS} = \sum_l \int \frac{d^2k}{(2\pi)^2} \varepsilon(\vec{k}) \psi_l^\dagger(\vec{k}) \psi_l^{\vphantom{\dagger}}(\vec{k}) + H_{\text{pert}},
    \label{eq:FS_Z2}
\end{equation}
where $\varepsilon(\vec{k})$ is the dispersion of the fermions, that vanishes at the Fermi surface specified by $\varepsilon(\vec{k})=0$, and $H_{\text{pert}}$ includes various possible perturbations, to be discussed in the following. We have assumed for simplicity that the gauged fermions, created by the operator $\psi^\dagger_l(\vec{k})$, do not carry any quantum number other than the layer index $l$ and momentum $\vec{k}$ (e.g., spin is not conserved).  

Before turning to the effects of inter-layer interactions, we first discuss stability of the Fermi surface in the limit where the layers are decoupled. It is not immediately obvious that the Fermi surface is stable, even in this limit, since the number of fermions is only conserved $\mod(2)$. Thus, pairing terms of the form $\Delta(\vec{k})\psi_l^\dagger(\vec{k}) \psi_l^\dagger(-\vec{k}) + \text{h.c.}$ are generally allowed in $H_{\text{pert}}$, and one may 
conclude that (except under fine-tuned circumstances\cite{zhang,arovas}) the Fermi surface is always gapped, with the exception of a discrete set of nodal points where both $\varepsilon(\vec{k})=0$ and $\Delta(\vec{k})=0$. Nevertheless, it turns out that stable Fermi surfaces are, in fact, possible in certain circumstances~\cite{Baskaran2009,Lai2011,Barkeshli2013,Hermanns2014}: (i) If both time reversal and inversion symmetries are broken, either spontaneously or explicitly, such that $\varepsilon(\vec{k})\ne \varepsilon(-\vec{k})$, then the Fermi surface remains even for a non-zero generic $\Delta(\vec{k})$, as long as the maximum of $|\Delta(\vec{k})|$ is below a certain critical value. (ii) Even in the presence of time reversal symmetry, there are situations where the Fermi surface is stable. This happens if the action of time reversal $\mathcal{T}$ on the gauged fermions is such that $\mathcal{T} \psi_l(\vec{k}) \mathcal{T}^{-1} = \psi_l(-\vec{k}+\vec{Q})$ with a certain non-zero wavevector $\vec{Q}$. In this case, $\varepsilon(\vec{k})\ne \varepsilon(-\vec{k})$ even in the presence of time reversal, and the Fermi surface is again stable\footnote{In this case, one can always choose a gauge where the action of time reversal is $\mathcal{T} \psi(\vec{k}) \mathcal{T}^{-1} = \psi(-\vec{k})$ by shifting the momentum of the fermions. However, in this case, the ``pairing term'' becomes modulated in space, as in a pair density wave (PDW) state, since includes pairs of electrons with a non-zero momentum. Pair density wave states are known to support stable Fermi surfaces~\cite{Baruch2008,Berg2009}.}. 

Note that the excitations near the Fermi surface are actually ``Bogoliubov-like'' quasiparticles, consisting of superpositions of $\psi_l$ and $\psi_l^\dagger$. Nevertheless, we can always diagonalize the quadratic part of the Hamiltonian including the pairing terms, bringing it to the form (\ref{eq:FS_Z2}) with only quartic and higher-order terms in $H_{\text{pert}}$.

For a generic dispersion $\varepsilon(\vec{k})$, short-range quartic terms, both intra-layer and inter-layer, are marginal by power counting, as in the RG analysis of a Fermi liquid~\cite{Shankar}. Higher-order terms are irrelevant. Taking into account the fact that the number of fermions must be conserved $\mod(2)$ in each layer, the most general quartic interaction is of the form
\begin{eqnarray}
    H_{\text{pert}}&=&\sum_{l,l'}\sum_{\{\eta_i=\pm 1\}}\int_{\{\vec{k}_i\}} \delta\left(\sum_{i=1}^{4}\eta_i\vec{k_i}\right) 
    \label{eq:Hpert}
    \\
    &\times& W_{l,l'}(\{\eta_i\},\{\vec{k}_i\}) \psi_l^{\eta_1}(\vec{k}_1) \psi_l^{\eta_2}(\vec{k}_2) \psi_{l'}^{\eta_3}(\vec{k}_3) \psi_{l'}^{\eta_4}(\vec{k}_4)\nonumber,
\end{eqnarray}
where we have introduced the notation $\psi_l^{\eta=+1} \equiv \psi_l$, $\psi_l^{\eta=-1} \equiv \psi^\dagger_l$. From the hermiticity of $H_{\text{int}}$, the interaction function must satisfy $W^{*}_{l,l'}(\{\eta_1\dots \eta_4\},\{\vec{k}_1\dots \vec{k}_4\}) = W_{l,l'}(\{-\eta_4\dots -\eta_1\},\{\vec{k}_4\dots \vec{k}_1\})$. 

The stability analysis of the Fermi surface including ``anomalous'' (
particle number non-conserving) terms of the form (\ref{eq:Hpert}) can be done along similar lines of the RG analysis for a Fermi liquid~\cite{Shankar}. We shall not go through the details of this analysis here, and only point out the main results. Most importantly, in the absence of a Kramers-like degeneracy, $\varepsilon(\vec{k})\ne \varepsilon(-\vec{k})$, the interaction function (\ref{eq:Hpert}) is exactly marginal. The imaginary part of the single-fermion self energy on the Fermi surface, to order $|W|^2$, is $
\Sigma''(\omega,T) \propto \max(\omega^2,T^2)\times  \log\left[\frac{\varepsilon_F}{\max(|\omega|,T)}\right]$ (where $\varepsilon_F$ is the Fermi energy), similar to that of an ordinary two-dimensional Fermi liquid. These results indicate that the resulting stable phase is a ``floating Fermi liquid'' with long-lived low energy fermionic quasiparticles that can propagate coherently within each layer, but not between layers. 

Note that the interaction (\ref{eq:Hpert}) includes terms that do not conserve the quasi-particle number $\psi_l^\dagger(\vec{k}) \psi_l^{\vphantom{\dagger}}(\vec{k})$ at a particular point on the Fermi surface. Such terms are not included in the standard Fermi liquid effective Hamiltonian, but nevertheless do not change its low-energy properties. They are analogous to the umklapp terms in a two-dimensional Fermi liquid, which are marginal under RG~\cite{Kumar1996} but do not destabilize the Fermi liquid.  

Since the layers are coupled by forward-scattering density-density interactions (included in Eq.~\ref{eq:Hpert}), the collective mode spectrum of the system may include ``zero sound'' modes that propagate in all three directions, depending on the nature of the inter-layer interactions. 

\subsubsection{Fermi surface with $U(1)$ gauge field}

Finally, we comment on the case of a layered system with a Fermi surface coupled to an emergent $U(1)$ gauge field in each layer. The Lagrangian density of an individual layer is given by
\begin{equation}
\mathcal{L}_l^{\mathrm{FS}} =\sum_{\alpha=1}^N \bar{\psi}_{i,l} \left[\partial_\tau + ia_{0,l}+ \varepsilon(-i\nabla + \vec{a}_l)\right] \psi_{i,l} + \frac{1}{4 e^2} f_{\mu \nu,l} f^{\mu \nu}_l.
\label{eq:FS_U1}
\end{equation}
As in the $\mathbb{Z}_2$ case, we must first address the stability of the Fermi surface in a single layer, before considering the effects of inter-layer coupling. Unfortunately, even this is at present an unsolved problem. The traditional expansion in $1/N$ turns out to be problemtaic, due to a proliferation of divergences in terms that are naively of high order in $1/N$, making the expansion unreliable at asymptotically low energies~\cite{SSLee,Metlitski2010,Holder2015}. 

A possible way to cure the problem~\cite{Nayak1994,Mross2010} is to generalize the bare ``photon dispersion'', coming from the last (Maxwell) term in Eq.~(\ref{eq:FS_U1}), to $\omega^2 \propto |\vec{q}|^{1+\epsilon}$. The physical case is $\epsilon=1$. The problem can be solved in a double expansion in both $1/N$ and $\epsilon$. To lowest order, one recovers the self-consistent one-loop approximation for the fermionic and gauge field self-energies~\cite{Altshuler1994,Mross2010}. Note that the price of this approach is that for $\epsilon<1$ the action is non-local in space, which entails potential subtleties in an RG analysis (which generally assumes a local form for the action, a property preserved by the RG transformation).

Proceeding nevertheless along these lines, Metlitski et al.~\cite{Metlitski2015} showed that the single-layer action (\ref{eq:FS_U1}) is stable in the presence of arbitrary quartic interactions, as long as they are sufficiently weak. In particular, the Fermi surface does not have a BCS instability towards Cooper pairing.\cite{sri} 

With the above caveats in mind, we may assess the effects of inter-layer interactions within the fixed point described by the $(\epsilon,1/N)$ expansion. 
The lowest-order gauge invariant inter-layer terms are of the form\footnote{For simplicity, we omit here some terms that are allowed by symmetry, e.g. a term of the form $(\nabla \times \vec{a}_l) (\nabla\times \vec{a}_{l'})$ (which is linearly independent from the term that appears at the end of Eq. \ref{eq:inter_FS}). These terms obey the same scaling as the term we have retained.}
\begin{equation}
    \mathcal{L}_{\text{inter}} = \sum_{l,l'} \left[u_{l,l'} n_l n_{l'} + g_{l,l'} f_{\mu \nu,l} f^{\mu \nu}_{l'} \right].
    \label{eq:inter_FS}
\end{equation}
Here, $n_l = \sum_i \psi_{i,l}^\dagger \psi_{i,l}$ is the density of fermions in layer $l$. The first term is irrelevant at the decoupled layer limit, by the same argument that makes the single-layer fixed point stable with respect to Cooper pairing. The second term is marginal by scaling, and renormalizes the dispersion of the (overdamped) photon mode, which now becomes dependent on the component of the wavevector perpendicular to the planes. Hence, at least within the $(\epsilon,1/N)$ expansion, a floating phase of Fermi liquids coupled to $U(1)$ gauge fields remains stable in the presence of inter-plane coupling. Whether this conclusion extends beyond this limit 
remains to be determined.

\section{Experimental diagnostic: Diverging conductivity anisotropy}\label{sec:experimental}

Having discussed the stability of floating phases, we now turn to possible experimental diagnostics of such phases.
The correlation function diagnostic in Sec~\ref{sec:fmop} is useful conceptually, but is 
{unlikely} to be measurable experimentally.
In this section, we discuss 
an experimentally accessible signature of a floating topological phase in the form of the electrical conductivity anisotropy at low temperatures $T\rightarrow 0$.

In a floating topological phase, it is possible that the transport coefficients
become \emph{parametrically} anisotropic in the limit $T\rightarrow0$.
In particular, the ratio of the in-plane and out of plane conductivites
diverges as $T\rightarrow0$. As far as we know, this is impossible
in ordinary, three dimensional (non-floating) phases, where we expect
the conductivities in all directions to scale in the same way as a
function of $T$, even when the microscopic parameters are strongly
anisotropic.

In gapless floating quantum spin liquids (QSLs), the anisotropy
of the thermal conducitivity has been studied by Werman et al~\cite{werman}.
However, in gapped spin liquids, phonons always dominate the thermal transport
at low temperatures, and the signatures of fractionalization of the
spin excitations are masked. Even in gapless QSLs, where the magnetic
excitations dominate at low temperatures, one may need to go to exceedingly
low $T$ to make the phonon contribution negligible. Here, we consider
the \emph{electrical} conductivity, which may be mediated by gapped ``holon'' {(or chargon)} excitations
(in both gapped and gappless QSLs). Since there is no phonon contribution
to the electrical conductivity, the electronic anisotropy may be more easily 
measurable.

In weakly disordered systems, we find conditions under which the anisotropy in the conductivity indeed diverges at low temperatures, in both gapped and gapless layered QSLs. 
The functional form of the divergence of the anisotropy 
depends on the type of QSL and the properties of its lowest-energy excitations.

Interestingly, if the charge carriers are localized and gapless, and hence the conductivity at low temperature occurs through variable-range hopping, it turns out that the resistivity anisotropy \emph{saturates}. Thus, a rapidly growing resistivity anisotropy can serve as a signature of floating topological phases, although in the presence of 
disorder, localization may ultimately preempt a true divergence.

We illustrate these principles by computing the resistivity anistropy in two examples: a gapped QSL and a gapless $\mathbb{Z}_2$ Dirac QSL, both with weak disorder. We then discuss possible non-perturbative effects of quenched disorder that may appear at low temperature, such as a modification of the low-energy spinon spectrum and Anderson localization.

\subsection{Gapped QSL}

Let us consider a layered, gapped $\mathbb{Z}_2$ QSL. The mechanism for conductivity at low temperature depends on the character of the lowest-energy charge excitations. Crucially, the transport of charge within the planes can be carried by fractionalized excitations (that carry a non-trivial gauge charge or flux), whereas inter-layer transport must be carried by gauge-neutral, charged excitations (such as electrons). This is the source of the parametric anisotropy in the limit $T\rightarrow 0$. 

As a concrete example, let us assume that the lowest-energy charged excitation in each layer is a holon that carries charge $+e$ and is ``electrically charged'' under the gauge field. The self-statistics of the lowest-energy holon excitation may either be fermionic or bosonic, depending on energetics.\cite{holon} Let us denote the holon creation operator by $h^\dagger_l(\vec{r})$. Then, we can define a ``spinon'' excitation $\psi_{l,\sigma}(\vec{r}) \sim h^\dagger_l(\vec{r}) c_{l,\sigma}^\dagger(r)$, where $c_{l,\sigma}^\dagger(r)$ creates an electron in layer $l$ with spin $\sigma=\uparrow,\downarrow$. The spinon carries spin $1/2$, zero electric charge, and a non-trivial gauge charge. Note that if the holon has fermionic self-statistics, then the spinon must have bosonic self-statistics and vice versa, such that the electron is a fermion. 

The low-energy effective Hamiltonian is then given by:
\begin{equation}
H=\sum_{\vec{k},l}\varepsilon_{\vec{k}}^{h}h_{\vec{k},l}^{\dagger}h^{\vphantom{\dagger}}_{\vec{k},l}+\sum_{\vec{k},\sigma,l}\varepsilon_{\vec{k}}^{\psi}\psi_{\vec{k},\sigma,l}^{\dagger}\psi^{\vphantom{\dagger}}_{\vec{k},\sigma,l}+H_{\text{dis}}+H_{\text{inter}},
\end{equation}
where $\vec{k}$ is the in-plane momentum, and at low energies we may expand $\varepsilon^h_{\vec{k}} = \Delta_h + \frac{k^2}{2m_h}+\dots$, with $\Delta_h$ and $m_h$ the holon gap and effective mass, respectively. Similarly, $\varepsilon^\psi_{\vec{k}} = \Delta_\psi + \frac{k^2}{2m_\psi} + \dots$.  
$H_{\text{dis}}$ and $H_{\text{inter}}$
are disorder and inter-plane coupling terms, to be discussed below. Crucially, we assume that the holon and spinon do not form an electron-like bound state. I.e., the gap for creating an electron excitation, $\Delta_c$, is larger than $\Delta_{\psi} + \Delta_h$. Under these conditions, both in-plane and out of plane conductivity are carried by spinons and holons, and electronic excitations can be ignored at low temperatures. We will discuss situations where this condition is violated below.  

Note that 
spinon number is only conserved mod 2 so terms of the form 
 $\psi^\dagger_{l,\uparrow}(\vec{r}) \psi^\dagger_{l,\downarrow}(\vec{r})$ are allowed in the Hamiltonian. However, such terms can be eliminated by a Bogoliubov transformation.  Quartic terms that describe interactions between spinons or between spinons and holons include, in general, terms that do not conserve the spinon number. Terms of the form $h^\dagger_l(\vec{r}) c^\dagger_{l,\sigma}(\vec{r}) \psi_{l,\sigma}(\vec{r})$ are also allowed. However, at low temperatures, excitations of all types are very dilute, and intra-layer interactions do not play an important role, as long as the spinon and holon interact repulsively and do not form a bound state.

To describe inter-layer transport, we include an inter-layer coupling of the form
\begin{equation}
    H_{\text{inter}} = g_\perp \int d^2 r \sum_{l,\sigma} h^\dagger_{l}(\vec{r}) h^{\vphantom{\dagger}}_{l+1}(\vec{r})
    \psi^\dagger_{l,\sigma}(\vec{r})
    \psi^{\vphantom{\dagger}}_{l+1,\sigma}(\vec{r}) + \text{h.c.}
    \label{eq:Hinter}
\end{equation}
In order to render the in-plane conductivity finite, we need to take into account terms that break translational symmetry in the plane. To this end, we include also a weak disorder potential $H_{\text{dis}}$, that may couple to both the spinons and holons. The explicit form of $H_{\text{int}}$ is not important for our present purpose. 
{We will assume that the disorder is sufficiently weak that its effects can be considered perturbatively - 
non-perturbative effects 
leading to localization } will be discussed in Sec.~\ref{sec:VRH}.  

The frequency-dependent conductivity perpendicular to the layers, $\sigma_\perp(\Omega)$, can be computed perturbatively in $g_\perp$. Assuming that the holons obey Bose statistics, whereas the spinons are fermions, $\sigma_\perp$ is given by
\begin{widetext}
\begin{equation}
\sigma_\perp(\Omega) = \frac{4\pi e^{2}|g_{\perp}|^{2}}{\hbar \Omega d}\int_{\vec{k},\vec{k}',\vec{q}}\left[\delta\left(\Omega-\varepsilon_{\vec{k}}^{h}+\varepsilon_{\vec{k}+\vec{q}}^{h}-\varepsilon_{\vec{k}'}^{\psi}+\varepsilon_{\vec{k}'-\vec{q}}^{\psi}\right)-\left(\Omega\rightarrow-\Omega\right)\right]
n_{B}\left(\varepsilon_{\vec{k}}^{h}\right)n_{B}\left(-\varepsilon_{\vec{k}+\vec{q}}^{h}\right)n_{F}\left(\varepsilon_{\vec{k}'}^{\psi}\right)n_{F}\left(-\varepsilon_{\vec{k}'-\vec{q}}^{\psi}\right).
\label{eq:sigmaperp}
\end{equation}
\end{widetext}
Here, $n_{F,B}(\varepsilon)$ are the Fermi and Bose functions, respectively, and $\int_{\vec{k}}\equiv \int\frac{d^2 k}{(2\pi)^2}$.

At low temperature, we may replace $n_{F,B}\approx e^{-\varepsilon/T}$. Taking the d.c. ($\Omega\rightarrow$ 0) limit, we obtain in a gapped QSL
\begin{align}
\begin{split}
\sigma_{\perp}	&= \frac{4\pi e^{2}|g_{\perp}|^{2}}{Td}\int_{\vec{k},\vec{k}',\vec{q}}\delta\left(\varepsilon_{\vec{k}}^{h}-\varepsilon_{\vec{k}+\vec{q}}^{h}+\varepsilon_{\vec{k}}^{\psi}-\varepsilon_{\vec{k}'-\vec{q}}^{\psi}\right)e^{-\frac{\varepsilon_{\vec{k}}^{h}+\varepsilon_{\vec{k}'}^{\psi}}{T}}
\nonumber\\
	&\propto e^{-\frac{\Delta_{h}+\Delta_{\psi}}{T}}.
\end{split}
\end{align}
In contrast, the in-plane conductivity is proportional to the density
of holons, and hence $\sigma_{\parallel}\propto e^{-\frac{\Delta_{h}}{T}}$
(with a prefactor proportional to the elsatic mean free time of a
holon in the plane). The anisotropy ratio 
\begin{equation}
\left(\frac{\sigma_{\parallel}}{\sigma_{\perp}}\right)_{\text{Gapped QSL}}\propto 
\exp\left[{\frac{\Delta_{\psi}}{T}}\right]
\end{equation}
 diverges at low $T$. 

Importantly, as noted above, we assumed that there is no attraction between spinons
and holons in a layer. 
Hence, the energy of an electron or a hole excitation is larger than $\Delta_h + \Delta_{\psi}$. 
If the minimal excitation energy of an electron, $\Delta_c$, is smaller than $\Delta_h+\Delta_{\psi}$, then the inter-plane conductivity is mostly carried by electronic excitations, 
so $\sigma_\perp \propto e^{-\Delta_c/T}$. 
Still, so long as $\Delta_c>\Delta_h$, the in-plane current is carried by holons and the anistropy still diverges as
$\sigma_\parallel/\sigma_\perp \propto \exp[(\Delta_c - \Delta_h)/T]$.  However,
if $\Delta_c<\Delta_h$, then both the in-plane and out of plane currents are carried by electrons, and the anisotropy ratio does not diverge in the 
$T\rightarrow0$ limit. Thus, a divergent anisotropy 
implies that the system
is in a floating phase, but the 
converse is not necessarily true: a gapped floating phase
may or may not have a divergent anisotropy.

\subsection{Gapless Dirac QSL}

Next, we consider the case of a gapless Dirac QSL coupled to a $\mathbb{Z}_2$ gauge field in each layer, considered in Sec.~\ref{sec:Z2Dirac}. The spinons are fermionic, with an in-plane action given by Eq.~(\ref{eq:Z2Dirac}), supplemented (assuming that the system is spin rotationally invariant) 
by a summation over the spin index. The system is electrically insulating, and the bosonic holons are assumed to be gapped with a minimal gap $\Delta_h$. The inter-layer coupling is taken to be of the form (\ref{eq:Hinter}). 

As in the gapped case, if the gap to create electron or hole excitations is smaller than $\Delta_h$, then the conductivity anisotropy ratio saturates at low temperature. In contrast, if the holon is the lowest-energy charged excitation, then the out of plane conductivity is of the form of Eq.~(\ref{eq:sigmaperp}), with the spinon dispersion given by $\varepsilon^{\psi}_{\vec{k}} = v|\vec{k}|$. We explore this case henceforth.

At low temperature, we approximate $n_B(\varepsilon^h_{\vec{k}})\approx e^{-\varepsilon^h_{\vec{k}}/T}$ and $n_B(-\varepsilon^h_{\vec{k}+\vec{q}})\approx -1$. The d.c. limit $\Omega\rightarrow 0$ is then taken. We may evaluate the integral in (\ref{eq:sigmaperp}) as follows. Due to the Fermi and Bose functions and the energy conservation condition, $|\vec{k}'|$
and $|\vec{q}|$ are both of order $\frac{T}{v}$, whereas $|\vec{k}|\sim\sqrt{2m_{h}T}$.
Therefore, $|\varepsilon_{\vec{k}}^{h}-\varepsilon_{\vec{k}+\vec{q}}^{h}|\approx\frac{\vec{k}\cdot\vec{q}}{m_{h}}\sim\sqrt{\frac{2T}{m_{h}}}\frac{T}{v}\ll T,$
and we may neglect the $\varepsilon_{\vec{k}}^{h}-\varepsilon_{\vec{k}+\vec{q}}^{h}$
term relative to $\varepsilon_{\vec{k}'}^{\psi}-\varepsilon_{\vec{k}'-\vec{q}}^{\psi}$
in the $\delta$ function. The integral over $\vec{k}$ can then be done independently of the integrals over $\vec{k}'$ and $\vec{q}$. This results in
\begin{equation}
    \sigma_\perp \propto T^3 e^{-\frac{\Delta_h}{T}}.
\end{equation}

The in-plane conductivity can be estimated from the Einstein relation:
$\sigma_{\parallel}=\kappa D$, where $\kappa$ is the compressibility
and $D$ is the diffusion constant. The compressibility is $\kappa\sim m_{h}e^{-\frac{\Delta_{h}}{T}}$. 
The diffusion constant is $D=\frac{1}{2}\langle v^{2}\rangle\tau$, where $\langle v^{2}\rangle\sim2T/m_{h}$
is the average velocity of the holons, and $\tau$ is the mean free
time of the holon due to impurity scattering (assumed to be temperature independent). 
We obtain 
\begin{equation}
\sigma_{\parallel}\sim Te^{-\frac{\Delta_{h}}{T}},
\end{equation}
and hence
\begin{equation}
\left(\frac{\sigma_{\parallel}}{\sigma_{\perp}}\right)_{\mathbb{Z}_2\text{ Dirac QSL}}\sim\frac{1}{T^{2}}.
\end{equation}
The conductivity anisotropy ratio diverges algebraically as $T\rightarrow 0$, unlike the gapped case where it diverges exponentially. As in the gapped case, if $\Delta_c < \Delta_h$ then the anisotropy ratio does not diverge, since both the in-plane and out of plane currents are carried by electrons.

\subsection{Non-perturbative disorder effects}
So far, we have treated the disorder potential in the layers as weak - basically, its only effect was to provide the holons with a finite in-plane transport lifetime. Here, we comment on possible non-perturbative effects of disorder, that couples either to the holons or the spinons. 
\label{sec:disorder}
\subsubsection{Spinon disorder}
In the case of a Dirac QSL, disorder may modify the nature of Dirac fermions qualitatively, depending on the symmetry of the problem. {In certain cases, such as in the Kitaev honeycomb model with time reversal symmetry, the disorder couples to the spinons as a random vector potential~\cite{Willans2010}. Then, the DOS scales as a non-universal power law with energy, with a power that depends continuously on the disorder strength~\cite{Ludwig}. The latter behavior may result in a non-universal power law dependence of the conductivity anisotropy ratio on temperature. Such behavior has been predicted for the thermal conductivity anisotropy in a layered Dirac QSL~\cite{werman}. 
If the disorder couples to the density of the Dirac spinons, then the density of states (DOS) approaches a non-zero constant at low energy~\cite{Fradkin1986}.}
We leave a detailed prediction of the conductivity anisotropy in this case, as well as a discussion of the anisotropy in the case of a QSL with a disordered spinon Fermi surface, to future studies. 

\subsubsection{Screw dislocations}
Interestingly, in a floating topological phase, the transport properties are qualitatively affected by the presence of lattice screw dislocations along the axis perpendicular to the layers. This is because such dislocations allow fractionalized excitations, e.g. spinons and holons in a layered QSL, to move coherently between layers. Thus, in the presence of a finite density of screw dislocations, the resistivity anisotropy saturates at sufficiently low temperature, with a saturation value that  depends on the density of screw dislocations. 

A screw dislocation in a layered system is a version of a ``twist defect'' discussed in multi-layered systems~\cite{Barkeshli2013}. Depending on the type of topological order in each layer, screw dislocations may also carry propagating gapless modes. Similar phenomena have been discussed in weak topological insulators~\cite{Ran2019}. If present, such gapless modes may further affect the resistivity anisotropy. 

\subsubsection{Localization and variable range hopping}
\label{sec:VRH}

Ignoring interactions, the motion of a low-energy holon is two-dimensional, and hence quenched disorder is generally expected to localize it. Moreover, for a generic disorder distribution, the holon excitations are then gapless (even if there is a finite gap in the absence of disorder), since at any given energy within the original gap there is 
a non-zero probability to find a localized state.  At finite temperature, the d.c. conductivity is then 
dominated by variable-range-hopping of either holons or electrons.  Moreover, as we shall now argue, at low enough $T$, 
the conductivity, even for currents in the in-plane direction, 
is always dominated by inter-plane hopping processes. 

To see this, consider hopping of charge between 
two localized states separated by a displacement $\vec R$.  The hopping rate, generically, depends exponentially on $\vec R$ as
$\nu \propto \exp[- S(\vec R)]$.  If the relevant charged excitations are fractionalized, then the behavior of $S(\vec{R})$ can be different depending on whether the two localized states are in the same plane or in different planes. For two localized states in the same plane,  $S=S_{\text{intra}}= |\vec R|/\xi_{\text{intra}}$ where   $\xi_{\text{intra}}$  is the in-plane localization length.  By contrast  for two localized  states in different planes, $S=S_{\text{inter}}=\sqrt{(R_\parallel/\xi_\parallel)^2 + (R_\perp/\xi_\perp)^2}$ where the subscripts $\parallel$  and  $\perp$  refer to  the in-plane and out-of-plane  directions, and $\xi_\parallel$ is not necessarily  equal to $\xi_{\text{intra}}$.   The important point is that for a given concentration of active states, $c$, the typical distance between neighboring states in a given plane grows in proportion to $c^{-1/2}$ while if interplane processes are considered it grows in proportion to $c^{-1/3}$.  

Thus, since  the concentration of  thermally accessible states vanishes as $T\to 0$, even in the limit $\xi_{\text{intra}}\gg \xi_\parallel > \xi_\perp$, at  low enough $T$ inter-plane hopping is always preferred. In this limit, the system can be considered as an anisotropoic continuum, with the result that the anisotropy has no effect on the $T$ dependence of the  resistivity (it has the usual form for variable range hopping) and the resistivity anisotropy, $\rho_\perp/\rho_\parallel = (\xi_{\parallel}/\xi_\perp)^2$.  Thus, while there could arise an intermediate $T$ regime in which the anisotropy in the hopping conductivity grows with decreasing $T$ (reflecting an intermediate regime in which hopping of fractionalized particles within a given plane dominates the transport in the $\parallel$ direction), at  low  enough  $T$ the anisotropy will always saturate.

\section{Concluding remarks}\label{sec:conclusion}

One of the more remarkable recent advances in our understanding of phases of matter is the discoveryof ``absolute stability''---the stability of a phase to absolutely any perturbation. The primary class of such absolutely stable phases are topological phases which exhibit topological order characterized by emergent gauge fields and fractionalization. Such phases are well known to be stable to all weak perturbations to their Hamiltonians\cite{wen1990ground}. Another class of absolutely stable phases which was discovered even more recently is that of discrete time crystals\cite{CvK2016} which are non-equilibrium phases of Floquet systems. Here the stability is with respect to arbitrary weak perturbations of the drive that preserve the period.

In the present paper we have examined the stability of topological phases to weak perturbations in dimensionality---which take place when we weakly coupled stacks of $d$ dimensional systems in the $(d+1)$st direction. Remarkably, as first observed by Senthil and Fisher\cite{SF} and studied at length in the present paper, topological phases are stable to such perturbations too, in the sense that the resulting ``floating'' phases are connected perturbatively to the strictly decoupled stacks of lower dimensional systems and share their universal properties. This stability is extraordinarily useful from the viewpoint of realizing topological phases in materials. While in cases of broken symmetry at long wavelength the higher, physical, dimension always wins and thus any interesting lower dimensional phenomena are always ultimately obscured, for topological phases such as spin liquids the opposite is true. Given that spin liquids are more likely to be found in two dimensions on account of stronger quantum fluctuations, it is highly encouraging that their weakly three dimensional continuations will---in a sharp sense---continue to exhibit the universal properties of the planar phases regardless of the details of the couplings in the third direction just as long as they are weak. In addition to this general observation we have presented a fairly general tool for identifying such floating spin liquids which is to study the low temperature anisotropy in their charge transport. We trust that this combination of the general and the particular will encourage and inform the ongoing search for spin liquids. 

Indeed, an analogous strategy for probing the presence of a candidate topological phase has recently been applied in the cuprate high temperature superconductors.  There, the observation\cite{taillefer} of an anomalously large non-electronic component of the in-plane thermal Hall conductivity, $\kappa_{xy}$, led to speculative interpretations\cite{subir,subir2,patrick}  in terms of a heat current carried by spinons associated with the presence, or near proximity of a chiral topological phase in the individual Cu-O planes. However, a followup study,\cite{taillefer2} found an effect of comparable magnitude in $\kappa_{xz}$, i.e. when the thermal current is in the direction perpendicular to the plane.  This observation is generally taken to rule out such an exotic explanation, and suggests instead that the thermal Hall response is associated with a still notable, and not fully understood aspect of the phonon dynamics.\cite{jingyuan,behnia}

Before concluding we briefly list some items worthy of further investigation:
\begin{itemize}
    \item {\it Frequency dependent conductivity:} We have focused on the anisotropy in the $T$ dependence of the conductivity. However, it may well be that in an insulator it is much more convenient to look at the variation of the electrical response with frequency.  {The optical conductivity of a gapless spin liquid is generically finite at frequencies below the charge gap\cite{Ng2007}}. The physics discussed in this paper will again imply that this response becomes infinitely anisotropic as $\omega \rightarrow 0$. The precise form of the anisotropy will require computation in various cases of interest.
    \item {\it Chiral phases:} Two dimensional spin liquids also come in a chiral variant where the long wavelength action contains a Chern-Simons term; indeed, there is strong numerical evidence that they can be found in relatively simple Hamiltonians. These phases, which break time reversal symmetry spontaneously, can be coupled three dimensionally in more interesting ways in which the topological order floats, while the discrete broken symmetry orders with various periods. Now we can end up with transport which proceeds either via the bulk or via a potentially gapless surface of the kind which has been studied in the context of layered quantum Hall systems. We note that the bulk response will have both longitudinal and transverse components of which the in-plane components ($\sigma_{xy}$ and $\kappa_{xy}$) will be non-trivial and especially interesting. We intend to return to this in future work. 
    The more general observation here is that such considerations will apply to other stacked phases where the lower dimensional phase exhibits both topological order and broken symmetries, such as quantum Hall ferromagnets. Indeed even stacked topological phases with unbroken symmetries---symmetry enriched topological phases---can potentially give rise to distinct floating phases wherein the same floating topological order is joined to different realizations of the now three dimensional symmetry group.
    \item {\it Metallic phases:} We have focused on insulating phases but the basic intuition also carries over to metallic phases that also exhibit fractionalization and emergent gauge fields, e.g. the ``orthogonal metal''~\cite{nandkishoreorthog,kaulorthog}. 
    A related phenomenon was explored in Ref. \onlinecite{Zhou2016} where the $T\to 0$ resistance anisotropy in a conventional (Fermi liquid) metallic phase was shown to diverge on approach to an assumed continuous quantum phase transition to a topological insulating phase with a spinon Fermi surface.
\end{itemize}

\section*{Acknowledgements } We thank T. Senthil, J. Tranquada, and V. Calvera for helpful comments.  SAK was supported, in part, by NSF grant \# DMR-1608055 at Stanford. EB was supported by the European Research Council (ERC) under grant HQMAT (grant no. 817799), by the US-Israel Binational Science Foundation (BSF), and by CRC 183 of the Deutsche Forschungsgemeinschaft.
TD acknowledges support from the Charlotte Elizabeth Procter Fellowship at Princeton University.

 \begin{appendix}
 \section{Layered gauge invariance from anisotropic lattice QED}\label{app:anisotropic_lgt}
 In this appendix, we argue that a model for an anisotropic lattice QED in $3+1$D (which has the usual gauge symmetry) can be described at low energy a model with a separate gauge symmetry on each layer such as those discussed in the main text.
 
Let us first describe the lattice $U(1)$ gauge theory~\cite{wilson}.
On each bond in of the $3+1$D hypercubic lattice, we define the $U(1)$ variable $U_\mu(\vec{x})=\exp i\int_{\vec{x}}^{\vec{x}+\hat{\mu}}A_\mu(\vec{x}^\prime) dx^\prime$, where $\mu=0,1,2,3$ corresponding to $\tau,x,y,z$. 
The anisotropic $U(1)$ gauge theory is described by the Euclidean action
\begin{eqnarray}
\begin{aligned}
S_{g} =& -\sum_{\vec{x},\mu,\nu<\mu}K_{\mu\nu} U_\mu(\vec{x})U_\nu(\vec{x}+\hat{\mu})U_\mu^\dagger(\vec{x}+\hat{\nu})U_\nu^\dagger(\vec{x}) + h.c.\\
\equiv& -\sum_{\vec{x},\mu,\nu<\mu} K_{\mu\nu} F_{\mu \nu}(\vec{x}) + h.c.
\end{aligned}
\end{eqnarray}
which is the discretization of $\cos(\mathrm{curl}A)$, with a direction-dependent coefficient $K_{\mu\nu}$ which encodes the anisotropy.  
The isotropic model $K_{\mu\nu}=K$ has a deconfined phase at weak coupling $K\ll1$, and a confined phase at strong coupling $K\gg 1$, with a phase transition in between.  
The anisotropic limit we wish to consider has $K_{\mu \nu}=K$ for $\mu,\nu<3$, and $K_{3 \mu}=K^\prime$, in the limit $K\gg q$ and $K^\prime \ll 1$.  
It was shown in Ref~\onlinecite{funielson} that in $3+1D$ in the pure gauge theory, this limit is always confined and therefore there is no stable layered phase (this reflects the fact that the $2+1$D $U(1)$ pure gauge theory is always confined).

We now discuss fermions, which for simplicity we take to be the naive discretization of $3+1$D (isotropic) Dirac fermions given by 
\begin{equation}
S_f = \frac{1}{2}\sum_{\vec{x},\mu,n} \bar{\psi}_n(\vec{x}) \gamma_\mu (U_\mu^\dagger(\vec{x})\psi_n(\vec{x}+\hat{\mu})-U_\mu(\vec{x}-\hat{\mu})\psi_n(\vec{x}-\hat{\mu}))
\end{equation}
where $\psi,\bar{\psi}=\psi^\dagger \gamma_0$ are four-component Grassmann spinors, $n=1,\dots,N$ label fermion flavors, and $\gamma_\mu$ satisfy $\{\gamma_\mu,\gamma_\nu\}=2\delta_{\mu\nu}$.
Note that if we ignore the gauge field by setting $U=1$, this actually leads to a low-energy description with $2^3 N$ Dirac fermions (due to fermion doubling).  
Since we are working at large $N$ (and are interested in a layered phase which will have Dirac fermions associated with each layer anyway), this does not pose a problem.

The full action is
\begin{equation}
S = S_g + S_f
\end{equation}
and has the gauge symmetry that involves sending
\begin{eqnarray}
\psi_n(\vec{x}) \rightarrow e^{i\alpha(\vec{x})} \psi_n(\vec{x})\\
\bar{\psi}_n(\vec{x}) \rightarrow e^{-i\alpha(\vec{x})} \bar{\psi}_n(\vec{x})\\
U_\mu(\vec{x}) \rightarrow e^{i(\alpha(\vec{x}+\hat{\mu})-\alpha(\vec{x}))}U_\mu(\vec{x})
\end{eqnarray}
We now argue that in the anisotropic limit $K^\prime \ll 1 \ll K$, this theory describes a gapless layered phase with a separate gauge invariance in each plane.

In the limit $K^\prime \ll 1$, $U_z$ is strongly fluctuating and can be integrated out explicitly via a ``high temperature'' expansion in $K^\prime$.
The effective action after integrating out $U_z$,
\begin{equation}
S_{\mathrm{eff}} = -\ln \int \mathcal{D} U_z e^{-S}
\end{equation}
can be obtained as a power series in $K^\prime$.
We use the convention
\begin{equation}
\int dU U^n = \delta_{n 0}
\end{equation}
for integrating over the $U(1)$ group element $U$.

Let us write $S_g = S_{g,\perp} + S_{g,z}$ and $S_f=S_{f,\perp}+S_{f,z}$ where $S_{g/f,z}$ involves terms with $U_z$, and $S_{g/f,\perp}$ are all other terms.  
Since $S_{g,z}$ is proportional to $K^\prime$, we may write $S=S_0+K^\prime S_1$ where $S_0=S_{g,\perp}+S_f$ and $S_1 = S_{g,z}/K^\prime$.
Then,
\begin{equation}
S_{\mathrm{eff}} = -\ln \int \mathcal{D} U_z e^{-S_0}(1-K^\prime S_1 + \frac{1}{2!}(K^\prime S_1)^2 - \dots)
\end{equation}
we see that the only terms which survive are those that contain no factors of any $U_z$.  
Furthermore, the logarithm means that only the connected terms survive.

At zeroth order in $K^\prime$, we have
\begin{align}
\begin{split}
S_{\mathrm{eff}}^{(0)} =& S_{g,\perp}+S_{f,\perp}\\
&+ \frac{1}{4}\sum_{\vec{x},n} (\bar{\psi}_n(\vec{x}) \gamma_3 \psi_n(\vec{x}+\hat{z}))(\bar{\psi}_n(\vec{x}+\hat{z})\gamma_3\psi_n(\vec{x}))
\end{split}
\end{align}
a density-density type interaction term between neighboring layers.  

At first order in $K^\prime$, we have 
\begin{align}
\begin{split}
S_{\mathrm{eff}}^{(1)} =& \frac{K^\prime}{4} \sum_{\vec{x},\mu<3,n,m} \bar{\psi}_n(\vec{x})\gamma_3 \psi_n(\vec{x}+\hat{z}) U^\dagger_\mu(\vec{x}) U_\mu(\vec{x}+\hat{z}) \\
&\times \bar{\psi}_m(\vec{x}+\hat{z}+\hat{\mu}) \gamma_3 \psi_m(\vec{x}+\hat{\mu})\\
&+ (\psi\leftrightarrow \bar{\psi}, U\leftrightarrow U^\dagger)
\end{split}
\end{align}
which has the form of a gauge-invariant nearest-neighbor pair hopping term involving two adjacent layers.

At second order, keeping only terms at fourth power in $\bar{\psi}, \psi$, 
\begin{align}
\begin{split}
S_{\mathrm{eff}}^{(2)} =& \frac{(K^\prime)^2}{4} \sum_{\langle x,x^\prime\rangle} \sum_{\mathrm{paths }\; C} 
\sum_{m,n}\bar{\psi}_n(\vec{x})\gamma_3 \psi_n(\vec{x}+\hat{z})\\
&\times \left(\prod_{x^{\prime\prime},\mu\in C}U^\dagger_\mu(\vec{x}^{\prime\prime}) U_\mu(\vec{x}^{\prime\prime}+\hat{z})\right) \\
&\times \bar{\psi}_m(\vec{x}^\prime+\hat{z}) \gamma_3 \psi_m(\vec{x}^\prime)\\
&+ (\psi\leftrightarrow \bar{\psi}, U\leftrightarrow U^\dagger)
\end{split}
\end{align}
where the first sum over $\langle x,x^\prime\rangle$ is over all next-nearest-neighbor pairs with the same $z$ coordinate, the second sum is over all paths $C$ of length $\mathrm{len}(C)=2$ connecting $x$ and $x^\prime$, and the product of $U$ is along the path connecting $x$ and $x^\prime$.
This has the form of a gauge-invariant next-nearest-neighbor pair hopping term.

At higher orders, we will have longer range pair hopping terms of this, made gauge-invariant by the Wilson line operators summed over all paths $C$.
At small $K^\prime$, these terms decay exponentially with distance as $(K^\prime)^{\mathrm{len}(C)}$, reflecting the fact that the $U_z$ degrees of freedom are gapped and strongly fluctuating.  

We may now take the continuum limit along $\tau$,$x$,$y$, in which the exponentially decaying term becomes essentially local.
The continuum model describes a stack of $U(1)$ gauge theories coupled to Dirac fermions with a separate gauge invariance on each layer, and local quartic couplings between layers.
As discussed in the main text, such quartic terms are irrelevant at low energies, for large $N$.
By construction, $S_{\mathrm{eff}}$ does not depend on $U_z$ and so therefore has a separate gauge invariance on each layer.  
The key result here is that the resulting theory is local;
had $K^\prime$ been too large, highly non-local terms would have been generated upon integrating out $U_z$.

 \end{appendix}

 \end{document}